\documentclass[twocolumn]{aastex62}

\usepackage{graphicx}
\usepackage{float}
\usepackage{hyperref}

\accepted{June 13, 2019}

\begin{document}

\title{Magnetic Flux Cancellation as the Trigger Mechanism of Solar Coronal Jets}

\email{ rmcglass@macalester.edu; panesar@lmsal.com}

\author[0000-0001-5258-1124]{Riley A. McGlasson}
\affiliation{Macalester College, Saint Paul, MN, USA}

\author[0000-0001-7620-362X]{Navdeep K. Panesar}
\affiliation{NASA Marshall Space Flight Center, Mail Code ST 13, Huntsville, AL 35812, USA}
\affiliation{Lockheed Martin Solar and Astrophysics Laboratory, 3251 Hanover Street, Bldg. 252, Palo Alto, CA 94304, USA}
\affiliation{Bay Area Environmental Research Institute, NASA Research Park, Moffett Field, CA 94035, USA}
\author{Alphonse C. Sterling}
\affiliation{NASA Marshall Space Flight Center, Mail Code ST 13, Huntsville, AL 35812, USA}
\author{Ronald L. Moore}
\affiliation{NASA Marshall Space Flight Center, Mail Code ST 13, Huntsville, AL 35812, USA}
\affiliation{Center for Space Plasma and Aeronomic Research (CSPAR), UAH, Huntsville, AL 35805, USA}


\begin{abstract}
Coronal jets are transient narrow features in the solar corona that originate from all regions of the solar disk: active regions, quiet sun, and coronal 
holes. Recent studies indicate that at least some coronal jets in quiet regions and 
coronal holes are driven by the eruption of a minifilament \citep{Sterling2015} following flux cancellation at a magnetic neutral line \citep{Panesar2016}. We have tested the veracity of that view by examining  60 random jets in quiet regions 
and coronal holes using multithermal (304 \AA, 171 \AA, 193 \AA, and 211 \AA) extreme 
ultraviolet (EUV) images from the Solar Dynamics Observatory (SDO)/Atmospheric 
Imaging Assembly (AIA) and line-of-sight magnetograms from the SDO/Helioseismic 
and Magnetic Imager (HMI). By examining the structure and changes in the magnetic field before, during, and after jet onset, we found that 85\%
of these jets resulted from a minifilament 
eruption triggered by flux cancellation at the neutral line. The 60 jets have a mean base diameter of 
$8800\pm3100$ km and a mean duration of $9\pm3.6$ minutes. These 
observations confirm that minifilament eruption is the driver and magnetic flux 
cancellation is the primary trigger mechanism for most coronal hole and 
quiet region coronal jets.
\end{abstract}


\section{Introduction} \label{sec:intro}
\begin{figure*}
    \centering
	\includegraphics[width=5in]{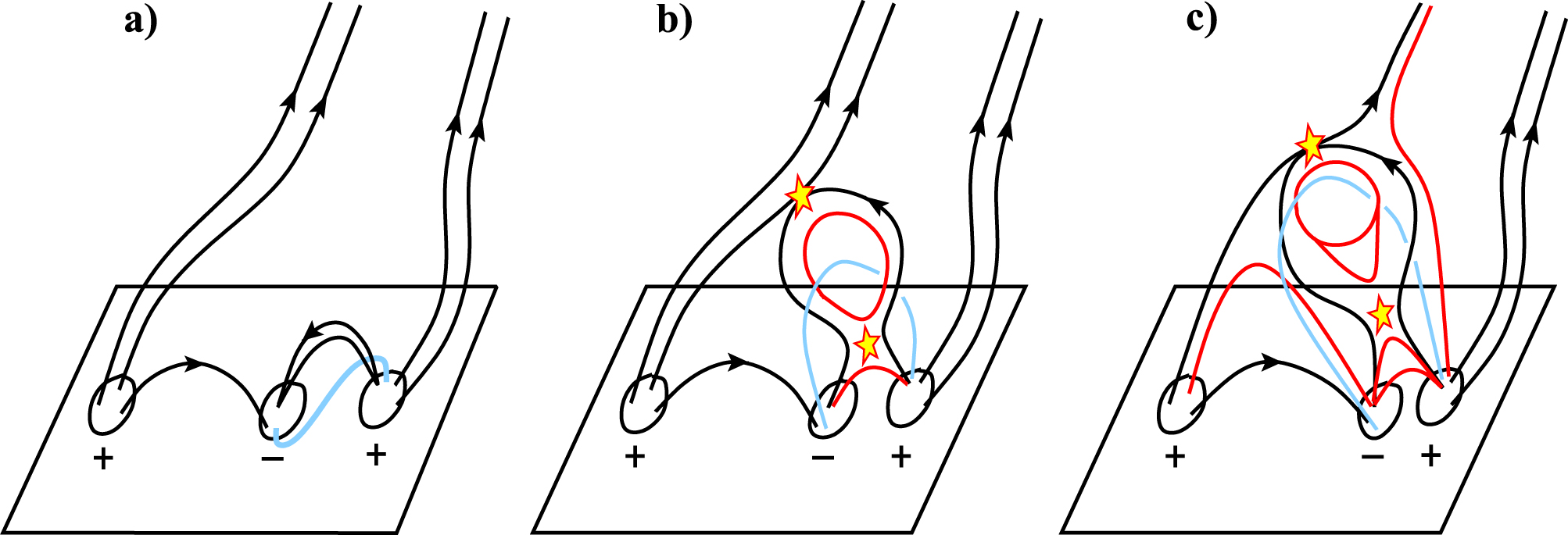}
	\caption{From \citet{Panesar2016}, an illustration of inferred coronal jet production. In this schematic, the solar surface is represented by the rectangular box, and the magnetic field lines are represented by the black lines. The ellipses marked by + and - represent positive-and negative-polarity flux patches. The blue line represents the minifilament, which in panel (a) is enveloped in  highly sheared and twisted magnetic field lines above the magnetic neutral line. As the negative minority flux patch approaches and cancels with the positive majority flux patch on the right, the minifilament field begins to erupt outwards, as in panel (b). This results in internal reconnection (lower star) between the legs of minifilament-carrying field, as well as external reconnection (upper star) of the minifilament-carrying field with adjacent outward-reaching field (with the two not necessarily starting at the same time). The external reconnection produces reconnected field represented by the two red lines external to the minifilament-carrying field in panel (c); one is the far-reaching red field line along which the jet plasma escapes, forming the jet spire, and the other is the red external loop below the external reconnection star.}
    \label{fig:jetfigRON}
\end{figure*}

Solar coronal jets are narrow, short-lived coronal features that occur frequently throughout the entire solar magnetic cycle \citep{Shimojo1998, Wang1998, Savcheva2007, Hong2011, Raouafi2016}. These events have been observed in all regions of the solar surface: in active regions \citep{Shibata1992, Innes2011,Panesar16a, Sterling2016, Sterling2017}, quiet regions \citep{Hong2011,Innes2016}, and coronal holes \citep{Cirtain2007, Nistico2009,Pucci2013,Sterling2015,Panesar2018}. In addition to being wide-spread across the solar surface, they also occur very frequently; \citet{Savcheva2007} found that jets in polar coronal holes occur at an average rate of 60 jets per day. Coronal jets are most often observed in extreme ultraviolet (EUV) \citep{Culhane2007, Zheng2013, Adams2014} and X-ray emission  \citep{Canfield1996,Yokoyama1998,Alexander1999,Cirtain2007,Glesener2012}.\par 


Coronal jets are characterized by a bright point (also known as the jet bright point, or JBP) at an edge of the base of the jet \citep{Shen2012,Young2014,Sterling2015,Panesar2016} and a jet spire of collimated outward moving plasma \citep{Shimojo1996,Shen2012, Young2014}. Additionally, recent studies \citep[e.g.][]{Hong2011,Shen2012, Sterling2015} have noted the presence of a minifilament that sits at the site of the forthcoming JBP and erupts to drive the jet. Minifilaments are usually seen to begin rising earlier ($\leq$1 minute) than the appearance of the JBP \citep{Moore2018}. 
Some earlier studies argued (based on observations with older data sets) that flux emergence may lead to jets \citep{Shibata1992, Shibata2007, Moreno-Insertis2008}. More recent on-disk jet studies have found in many cases consistent evidence of magnetic flux cancellation prior to and during jet onset \citep{Hong2011,Huang2012,Young2014,Panesar2016}. \citet{Panesar2016} investigated the magnetic trigger mechanism of 10 randomly-selected on-disk quiet region jets and found that flux cancellation was the trigger of each of these jets. In addition to larger coronal jets, small-scale jets \citep{Panesar2018b} are also observed to  erupt from neutral lines at which flux is canceling.  As shown in \citet{Panesar2017}, jets typically occur following the formation of a cool-minifilament-plasma residing over the magnetic neutral line between canceling positive-polarity and negative-polarity flux. Thus, coronal jets appear to be miniature versions of larger-scale eruptions that make coronal mass ejections (CMEs; \citealt{Sterling2018}).  \par

Figure \ref{fig:jetfigRON} is a diagram of the inferred jet-production process from \citet{Panesar2016}, based on their AIA and HMI observations. Figure \ref{fig:jetfigRON}(a) shows schematically two neighboring bipoles having adjacent negative-polarity ends on the photosphere, the smaller of the two containing a highly sheared and twisted field of a flux rope that holds a minifilament. The minifilament sits directly over the magnetic neutral line between the positive- and negative-polarity flux patches. As the opposite polarity flux patches continue to approach each other and cancel (Fig. \ref{fig:jetfigRON} b), the minifilament field becomes increasingly unstable due to unleashing of the flux rope through slow tether-cutting  flux cancellation driven by the converging photospheric flows. The magnetic pressure in the increasingly unleashed flux rope pushes the minifilament upwards, and the field starts to erupt. This results in runaway tether-cutting internal reconnection \citep{moore92} (at a  current sheet) within the legs of the field enveloping the erupting minifilament flux rope. As this field erupts outwards it reconnects on its outside with the neighboring far-reaching oppositely-directed field lines (``external reconnection"). The low lying red loop in Fig. \ref{fig:jetfigRON}(b) represents the JBP that appears at the location where the minifilament originally resided. The external reconnection creates new magnetic connections in two places: closed field lines over the large, left-most bipole (Fig. \ref{fig:jetfigRON} c) that causes external brightenings to appear in and around the base of the jet in EUV images,  and reconnected open or far-reaching field lines (Fig. \ref{fig:jetfigRON} c). Minifilament plasma escapes out along the reconnected open or far-reaching field to form part of the jet spire.\par

The above mentioned papers have each analyzed only one event or a limited number of jet events, so the triggering and driving mechanisms of coronal jets are not yet fully established. In this paper, we test  with a larger sample of non-active-region coronal jets whether flux cancellation and minifilament eruption usually cause coronal jets. To this end, we look for minifilament eruptions in and analyze the magnetic field evolution of 60 randomly selected solar coronal jets in quiet regions and coronal holes. We use SDO/AIA EUV movies to follow the jet emission, and use SDO/HMI line-of-sight magnetograms to track the magnetic flux over time at the jet-base region. With the high resolution and high cadence HMI magnetograms we are able to see flux cancellation that might not have been visible in other studies that used magnetograms with lower quality. Our study of 60 solar coronal jets indicates that minifilament eruption is usually the driver and flux cancellation is usually the trigger mechanism of solar coronal jets.



\startlongtable
\begin{deluxetable*}{c c c c c c c r@{$\pm$}l r@{$\pm$}l}
	\tablewidth{0pt}
	\tabletypesize{\footnotesize}
	\tablenum{1}
	\tablecaption{Measured Parameters of Coronal Jet Observations}
	
	\tablehead{
			\colhead{Event} & 
		\colhead{Date} & 
		\colhead{Time\textsuperscript{a}} & 
		\colhead{Location\textsuperscript{b}} & 
		\colhead{Region\textsuperscript{c}} & 
		\colhead{Erupting MF} &
		\colhead{Visibility of} & 
		\multicolumn{2}{c}{Jet Duration\textsuperscript{f}} & 
		\multicolumn{2}{c}{Base Width\textsuperscript{g}} 
		\\\colhead{No.}&
	    \colhead{} & 
		\colhead{(UT)}&
		\colhead{x,y (arcsec)}&
		\colhead{}&
		\colhead{Visibility\textsuperscript{d}}&
		\colhead{Cancellation\textsuperscript{e}}& 
		\multicolumn{2}{c}{(minutes)}& 
		\multicolumn{2}{c}{(km)}}
	
	\startdata
1 &	2017-05-18 & 11:40 & -600,64 & qr & y & y & \hspace{.55cm}13&3 & 7500 &  1900 \\
2 &	2017-04-16 & 6:51 & -73,321 & qr & y & y & 9 &  2 & 8000 &  300 \\
3 &	2017-04-14 & 9:04 & 82,-209 & qr & y & y & 12 &  2 & 6400 &  700 \\
4 &	2017-04-11 & 23:38 & 420,475 & qr & y & y & 12 &  2 & 8000 &  1800 \\
5 &	2017-04-10 & 16:56 & 199,127 & qr & y & y & 10 &  1 & 8000 &  200 \\
6 &	2017-04-04 & 8:29 & -496,75 & qr & y & y & 7 &  2 & 5700 &  100 \\
7 &	2017-03-27 & 3:14 & 406,-15 & ch & y & y & 13 &  2 & 6700 &  1800 \\
8 &	2017-03-24 & 2:15 & -211,-135 & ch & y & y & 7 &  2 & 12600 &  2600 \\
9 &	2017-03-24 & 5:48 & -180,-134 & ch & y & y & 5 &  2 & 11800 &  500 \\
10&	2017-03-23 & 0:35 & 533,-122 & ch & y & y & 8 &  2 & 5100 &  700 \\
11&	2017-03-22 & 4:56 & -485,69 & ch & y & y & 15 &  2 & 8600 &  170 \\
12&	2017-03-21 & 6:31 & 245,244 & ch & y & ** & 16 &  2 & 11700 &  900 \\
13&	2017-03-18 & 0:53 & -480,-82 & ch & y & y & 10 &  2 & 5000 &  1400 \\
14&	2017-03-17 & 7:00 & -640,-145 & ch & y & y & 4 &  1 & 9000 &  900 \\
15&	2017-03-17 & 6:54 & -686,-175 & qr & y & ** & 5 &  1 & 9900 &  2350 \\
16&	2017-03-13 & 9:49 & 491,261 & qr & $\dagger$ & y & 5 &  3 & 6340 &  800 \\
17&	2017-03-10 & 6:15 & -66,140 & qr & y & ** & 8 &  1 & 5000 &  750 \\
18&	2017-03-09 & 0:17 & -62,-114 & qr & y & y & 9 &  2 & 7100 &  1100 \\
19&	2017-03-05 & 3:06 & 372,208 & qr & y & y & 8&  4 & 4600 &  1300 \\
20&	2017-03-01 & 1:50 & -305,252 & qr & y & y & 14 &  2 & 3900 &  2000 \\
21&	2017-02-28 & 21:53 & 659,91 & ch & y & ** & 6 &  2 & 7450 &  1100 \\
22&	2017-02-28 & 18:23 & 659,91 & ch & y & y & 9 &  3 & 12839 &  3848 \\
23&	2017-02-28 & 21:47 & 694,239 & qr & $\dagger$ & y & 6 &  1 & 3490 &  200 \\
24&	2017-02-22 & 15:40 & -230,32 & qr & y & y & 8 &  2 & 6300 &  1300 \\
25&	2017-02-20 & 4:57 & -256,247 & ch & y & y & 6 &  2 & 7200 &  150 \\
26&	2017-02-14 & 5:29 & -269,-338 & ch & y & y & 11 &  2 & 6000 &  500 \\
27&	2017-02-02 & 23:14 & 647,-439 & ch & y & y & 7 &  2 & 5900 &  1400 \\
28&	2017-01-31 & 1:15 & 214,-49 & ch & y & y & 6 &  2 & 4000 &  1300 \\
29&	2017-01-29 & 15:09 & -134,-353 & ch & y & y & 18 &  3 & 4700 &  550 \\
30&	2017-01-27 & 15:25 & -421,-284 & ch & y & y & 6 &  1 & 7200 &  150 \\
31&	2017-01-24 & 9:38 & 530,-92 & qr & y & y & 13 &  1 & 7000 &  400 \\
32&	2017-01-23 & 1:34 & -435,-3 & ch & y & y & 12 &  2 & 7400 &  1300 \\
33&	2017-01-22 & 13:22 & -734,-137 & qr & y & y & 11 &  2 & 5700 &  1200 \\
34&	2017-01-19 & 22:58 & -49,83 & ch & y & y & 11 &  2 & 4800 &  500 \\
35&	2017-01-17 & 11:54 & 279,510 & qr & y & ** & 4 &  1 & 7300 &  250 \\
36&	2017-01-09 & 5:10 & 155,402 & qr & y & y & 7 &  2 & 7800 &  2600 \\
37&	2017-01-07 & 14:43 & -373,403 & qr & y & ** & 15 &  2 & 4500 &  250 \\
38&	2017-01-03 & 18:06 & -31,-660 & ch & y & ** & 9 &  3 & 6400 &  650 \\
39&	2017-01-02 & 16:09 & -114,-58 & ch & y & y & 11 &  3 & 10100 &  3500 \\
40&	2016-12-31 & 8:20 & 654,-223 & qr & y & y & 15 &  2 & 12400 &  4000 \\
41&	2016-12-28 & 19:53 & -50,176 & ch & y & y & 7 &  2 & 7100 &  1500 \\
42&	2016-12-28 & 0:20 & 398,79 & qr & y & ** & 6 &  3 & 17700 &  2200 \\
43&	2016-12-27 & 10:39 & -462,-385 & qr & y & y & 5 &  2 & 11100 &  500 \\
44&	2016-12-26 & 16:30 & -532,29 & ch & y & y & 8 &  2 & 4600 &  500 \\
45&	2016-12-11 & 5:33 & -304,429 & qr & y & y & 6 &  2 & 6400 &  650 \\
46&	2016-12-11 & 5:00 & -390,405 & qr & $\dagger$ & y & 6 &  2 & 5000 &  1500 \\
47&	2016-12-08 & 2:22 & -168,303 & ch & y & y* & 7 &  2 & 9412 &  724 \\
48&	2016-12-03 & 3:53 & -470,-476 & ch & $\dagger$ & y & 11 &  3 & 6400 &  750 \\
49&	2016-12-03 & 4:20 & -358,-489 & ch & y & y & 17 &  2 & 11700 &  500 \\
50&	2016-11-28 & 16:16 & 7,-83 & qr & y & y & 9 &  2 & 6800 &  300 \\
51&	2016-11-25 & 7:09 & -375,-25 & qr & y & y & 4 &  2 & 6400 &  1500 \\
52&	2016-11-20 & 12:26 & -57,433 & ch & y & y & 15 &  2 & 8600 &  400 \\
53&	2016-11-19 & 9:12 & -41,-511 & qr & y & y & 13 &  2 & 12200 &  1800 \\
54&	2016-10-24 & 5:48 & -14,413 & ch & y & y* & 6 &  3 & 6700 &  1300 \\
55&	2016-09-24 & 6:49 & -661,-144 & ch & y & y* & 10 &  2 & 8300 &  1000 \\
56&	2016-09-22 & 13:26 & -663,354 & ch & y & ** & 4 &  3 & 8400 &  1100 \\
57&	2016-09-18 & 16:10 & 75,103 & qr & y & y & 7 &  3 & 8250 &  800 \\
58&	2016-09-17 & 2:08 & -105,-234 & ch & y & y & 7 &  3 & 10200 &  1400 \\
59&	2016-09-12 & 10:15 & -167,312 & qr & y & y & 8 &  1 & 9700 &  1300 \\
60&	2016-09-11 & 11:28 & -393,268 & qr & y & y & 10 &  2 & 6900 &  1200 \\
\noalign{\smallskip}\tableline\tableline \noalign{\smallskip}
average&	 &  &  &  & 90\% & 85\% & 9 &  4 & 8800 &  3100 \\
\enddata
	
	\singlespace
	\tablecomments{
		\\\textsuperscript{a}The approximate time of jet spire formation.\\\textsuperscript{b}Location of the jet on the solar disk.\\\textsuperscript{c} The region, coronal hole (ch) or quiet region (qr), in which the jet resides.\\\textsuperscript{d}y indicates the presence of a visible erupting minifilament (MF) during the jet.\\\textsuperscript{e}y indicates the presence of visible cancellation prior to jet onset.\\\textsuperscript{f}This is the duration of the spire's growth, measured from start of the spire to maximum spire length. Uncertainties are estimated from 3 or 4 repeated measurements along the spire.\\\textsuperscript{g}Width of jet base region measured at its widest point approximately 1 minute prior to jet formation.  Uncertainties are estimated from from 3 or 4 repeated measurements (3 to 4) of the base width.\\\textsuperscript{*}Some emergence prior to cancellation.\\\textsuperscript{**}Convergence not clear.\\\textsuperscript{$\dagger$}Minifilament not clear.}
	\label{tab:jetmeasurements}
\end{deluxetable*}

\section{Instrumentation and Data} \label{sec:Inst/Data}
This study uses multithermal (171 \AA, 193 \AA, 211 \AA, and 304 \AA) EUV images from the Solar Dynamics Observatory (SDO)/Atmospheric Imaging Assembly (AIA) to study the eruption of the minifilament and jet spire. SDO/AIA takes high-resolution (0\arcsec.6 pixel$^{-1}$) and high temporal cadence (12s) full-Sun images in seven EUV wavelengths \citep{Lemen2012}. We primarily used 171 \AA{} and 193 \AA{} images for this study, as we found pre-jet minifilaments to be best seen in 171 \AA{} (600,000 K, \citealt{Lemen2012}) and coronal hole jets to be easily seen in 193 \AA{}  (peak temperatures of 1,500,000 K, \citealt{Lemen2012}). We used both 193 \AA{} data and 211 \AA{} (peak temperatures of 2,000,000 K) data to confirm the presence of a jet in coronal holes and the quiet Sun, but used 171 \AA{} data to measure base width and jet lifetime.\par

To analyze the magnetic field evolution, we used line-of-sight magnetograms from the SDO/Helioseismic and Magnetic Imager (HMI). SDO/HMI produces high-resolution (0\arcsec.5 pixel$^{-1}$ ) magnetograms with a temporal cadence of 45s that allow us to  examine closely the photospheric magnetic field around the jet-base region \citep{Schou2012}. We use magnetogram data to track the evolution of the photospheric magnetic flux in the jet-base region from approximately 6 hours prior to jet onset to approximately 1 hour after the jet.\par

For this study we used JHelioviewer  \citep{jhelioviewer} to randomly find 60  coronal jets in zoomed-in quiet regions and zoomed-in coronal holes, 30 jets in each type of region. We downloaded SDO/AIA and SDO/HMI for 200\arcsec $\times$ 150\arcsec areas surrounding each of the 60 jets from the Joint Science Operations Center cutout service\footnote{\url{http://jsoc.stanford.edu/ajax/lookdata.html?ds}} (JSOC). We downloaded AIA data at a 12-second cadence and HMI data at a 45 second cadence to observe the structure and magnetic evolution of each jet.  All of the data were derotated to the middle time of each data set in order to remove the drift from solar rotation. With the derotated data, we selected a smaller field of view focused closely on the jet-base region to perform measurements and to make movies of each jet. All jets and their parameters can be found in table \ref{tab:jetmeasurements}.\par

\begin{figure*}[t]
    \centering
	\includegraphics[width=6in]{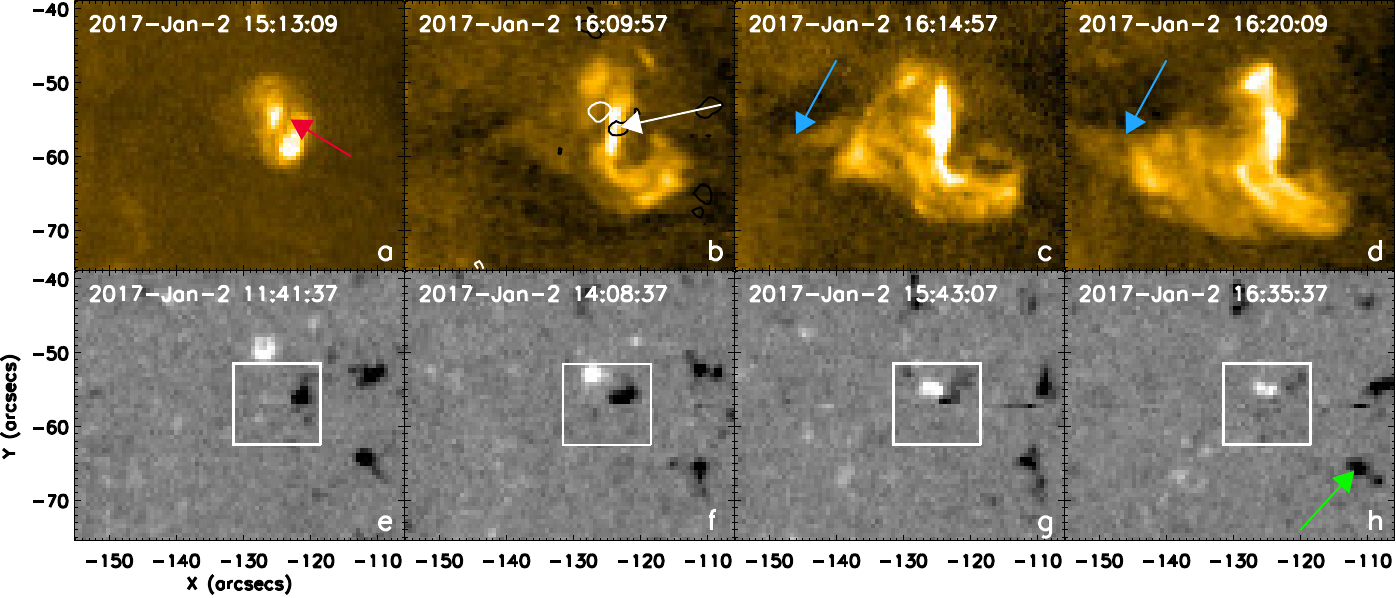}
	\caption{An example coronal hole jet; observed on 2017 January 2  (Event 39). Panels (a)-(d) show 171 \AA{} AIA intensity images. The red arrow in panel (a) points to the pre-eruption minifilament. The white arrow in panel (b) shows the JBP. The blue arrows in panels (c and d) indicate the jet spire. White and black contours in (b) outline the positive-and negative-polarity flux patches, respectively, at the time of the AIA image. Panels (e)-(h) show HMI magnetogram images of the same region. The boxed areas in panels (e)-(h) enclose the measured negative flux plotted in Figure \ref{fig:Jan2plot}.  The green arrow in panel (h) points to a nearby patch of majority (negative) flux to which EUV loops extend in (c) and (d).  We have included movies in the online version of this paper showing the complete evolution of this jet in AIA 171 \AA{} (MOVIE1) as well as in the corresponding HMI magnetogram data (MOVIE2).}
	\label{fig:Jan2image}
\end{figure*}
\section{Results} \label{sec:results}

\subsection{Overview} \label{sec:overview}
We report on the structure and magnetic origins of 60 EUV coronal jets $\lesssim$ 50 degree from disk center, 30 of which are from coronal holes, and 30 from quiet regions (see Table \ref*{tab:jetmeasurements}). We looked for erupting minifilaments at the times of the jets in these images. By registering the EUV images from SDO/AIA with the magnetograms from SDO/HMI, we were able to track the flux changes that occurred leading up to these jets.  Table \ref{tab:jetmeasurements} lists the 60 jets and their measurements. In sections \ref{sec:CH} and \ref{sec:QR} we present two examples of coronal hole jets and  two examples of quiet region jets. \par

\begin{figure}[h]
	\plotone{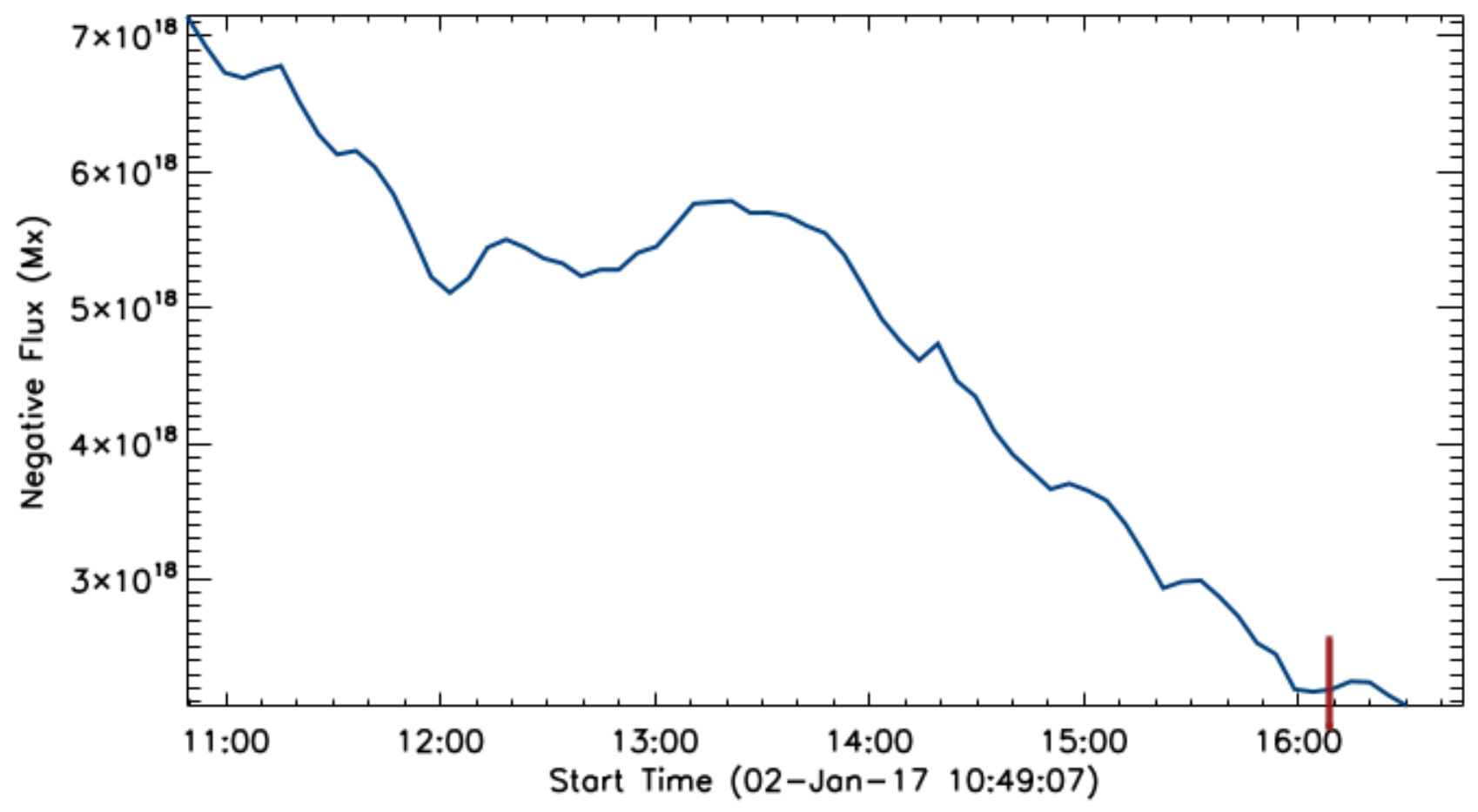}
	\caption{Flux-time plot from the base of the coronal hole jet of Figure \ref{fig:Jan2image}. This shows the majority (negative) flux integrated over the box in Figure \ref{fig:Jan2image}(e)-(h) as a function of time. The red line designates the time of jet onset (16:09 UT).}
	\label{fig:Jan2plot}
\end{figure}

To examine the magnetic field evolution quantitatively, we measured either the majority-polarity or the minority-polarity flux, whichever was the more isolated in the jet base region from prior to ($\sim6$ hours) until well after ($\sim1$ hour) the jet onset. We measure the flux of that polarity in a box that encloses the isolated flux patch and has no discernible flow of flux of that polarity across its perimeter. Figures \ref{fig:Jan2plot}, \ref{fig:Mar22plot}, \ref{fig:Nov25plot}, and \ref{fig:April10plot} show the measured flux plotted as a function of time. For each case, we integrated the isolated-polarity flux over the entire box at each time step.

We also found the base width and jet duration  for each jet (see Table \ref{tab:jetmeasurements}). 
To find the base width of each jet, we measured the longest side of the jet base approximately one minute prior to the onset of the jet spire in each event. For the jet duration, we measure the time from when the spire is first discernible until it reaches its maximum length; i.e., 	by “duration” we mean the growth time of the spire.  Uncertainties in the base width and	durations were estimated as the standard deviation from repeated measurements of the 	quantities (three or four times for each measurement). We ignored line-of-sight 	foreshortening effects. Because of this effect, some of the base-width measurements could be underestimated; in the most extreme case the underestimate factor would be $\sim$1/cos(50), about 50\%; but most events would be effected less than this, in part because they are much less than 50 deg from disk center. All of these measurements were made using 171 \AA{} EUV images.

\subsection{Coronal Hole Jets} \label{sec:CH}
	We randomly selected 30 EUV coronal hole jets from September 2016 to May 2017 (see Table \ref{tab:jetmeasurements}). We present two detailed examples of these coronal hole jets in sections \ref{sec:Jan2} and \ref{sec:Mar22}. 

\subsubsection{Coronal Hole Jet on January 2, 2017 (Event 39)} \label{sec:Jan2}
\begin{figure*}[t]
    \centering
	\includegraphics[width=6in]{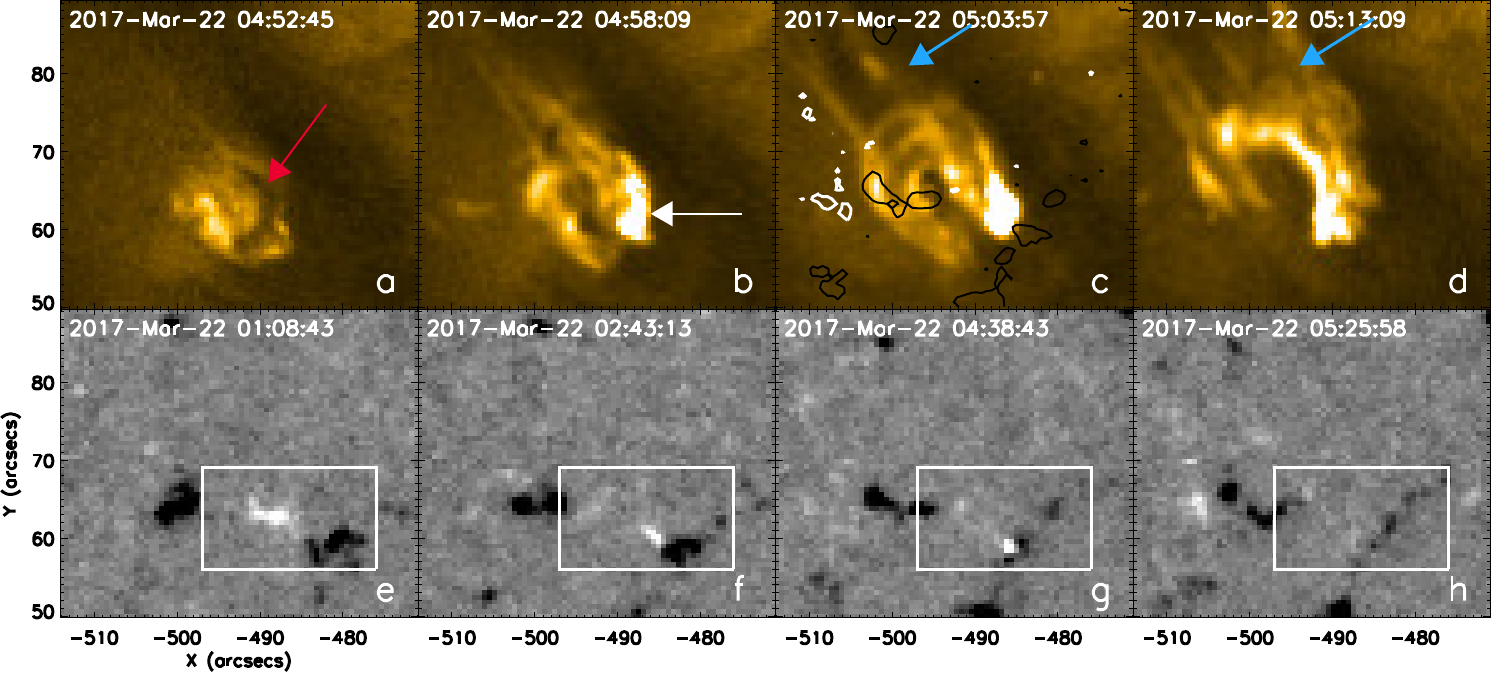}
	\caption{Another example coronal hole jet; observed on 2017 March 22  (Event 11). Panels (a)-(d) show 171 \AA{} AIA intensity images. The red arrow in panel (a) points to the erupting minifilament. The white arrow in panel (b) shows the JBP. The blue arrows in panels (c and d) indicate the jet spire, which includes the erupting minifilament. Panels (e)-(h) show HMI magnetogram images of the same region. White and black contours in (c) represent the positive and negative polarity flux, respectively, at the time of the AIA image. The boxed areas in panels (e)-(h) enclose the measured positive flux plotted in Figure \ref{fig:Mar22plot}. Movies showing the entire evolution of this jet are available in the online version of this paper in AIA 171 \AA{} (MOVIE3) as well as in the corresponding HMI magnetogram data (MOVIE4).}
	\label{fig:Mar22image}
\end{figure*}
A coronal hole jet was observed starting at approximately 16:09 on January 2, 2017 (see Table \ref{tab:jetmeasurements} for details). A pre-jet minifilament was present in the jet base region (Fig. \ref{fig:Jan2image}(a), red arrow; MOVIE1, available in the online version of this paper) residing above the magnetic neutral line between majority (black) and minority (white) flux patches (Fig. \ref{fig:Jan2image}(b,f)). This minifilament is visible in Fig. \ref{fig:Jan2image}(a), almost an hour prior to the jet. The JBP (Fig. \ref{fig:Jan2image}(b)) then appeared at the site of the minifilament as the minifilament slowly began to rise, and then erupted into the jet spire (Fig. \ref{fig:Jan2image}(c,d)). This progression appears analogous to that of typical filament eruptions, where the flare arcade grows over the neutral line in the wake of  the erupting filament. The JBP is a miniature flare arcade made by internal reconnecton  of the legs of the erupting minifilament field.
The duration  of the jet spire was $11\pm3$ minutes, and the jet had a base width of approximately 12,600 km. We also observed new loop brightenings extending to majority flux patches (e.g., Fig. \ref{fig:Jan2image}(d), green arrow in Fig. \ref{fig:Jan2image}(h)).\par

Figure \ref{fig:Jan2image}(e)-(h) shows the HMI magnetogram data for the jet observed on 2017 January 2. Initially, about 5 hours prior to the jet, the positive (white) and negative (black) flux patches were far apart (MOVIE2, available in the online version of this paper). With time, the flux patches approached each other and eventually  started canceling at the magnetic neutral line between them. This flux cancellation occurred continuously for 5 hours, until the minifilament erupted, producing the jet. We measured the isolated negative majority flux for 5.5 hours in the region indicated by the white box in Figure \ref{fig:Jan2image} panels (e)-(h). Figure \ref{fig:Jan2plot}  shows the total negative flux in the box as a function of time. It shows an overall decrease over the 5 hours prior to the jet, providing clear evidence of flux cancellation in the jet base region leading to the eruption of the minifilament and jet onset (red line in Fig. \ref{fig:Jan2plot}). 
This flux evolution is not monotonic, however, as  there is hump in the curve between $\sim$13:00-14:00 UT. Inspection of magnetograms over this period indicates that the increase is due to coalescence of negative flux in the jet-base region over this time (between Figs. 2(e) and 2(f)), resulting in weak negative  flux  rising above the noise level.  Even so, there is still an overall downward trend in the negative-polarity flux plot, consistent with cancellation dominating the field changes over the period.\par


\subsubsection{Coronal Hole Jet on March 22, 2017 (Event 11)} \label{sec:Mar22}

\begin{figure}[h!]
	\plotone{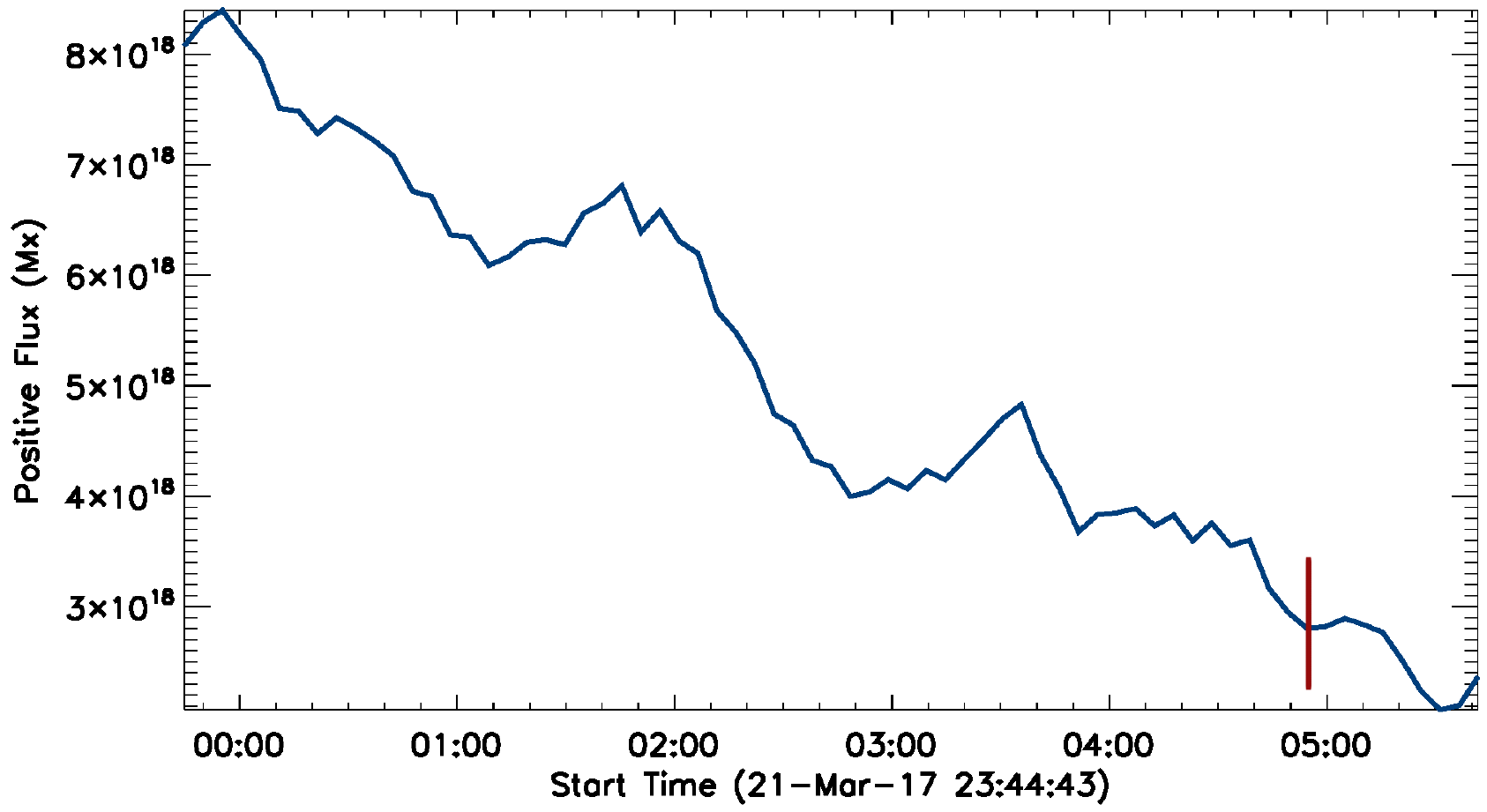}
	\caption{Flux-time plot from the base of the coronal hole jet of Figure \ref{fig:Mar22image}. This shows the minority (positive) flux integrated over the box in Figure \ref{fig:Mar22image}(e)-(h) as a function of time. The red line designates the time of jet onset (04:56 UT).}
	\label{fig:Mar22plot}
\end{figure}

\begin{figure*}[t]
    \centering
	\includegraphics[width=6in]{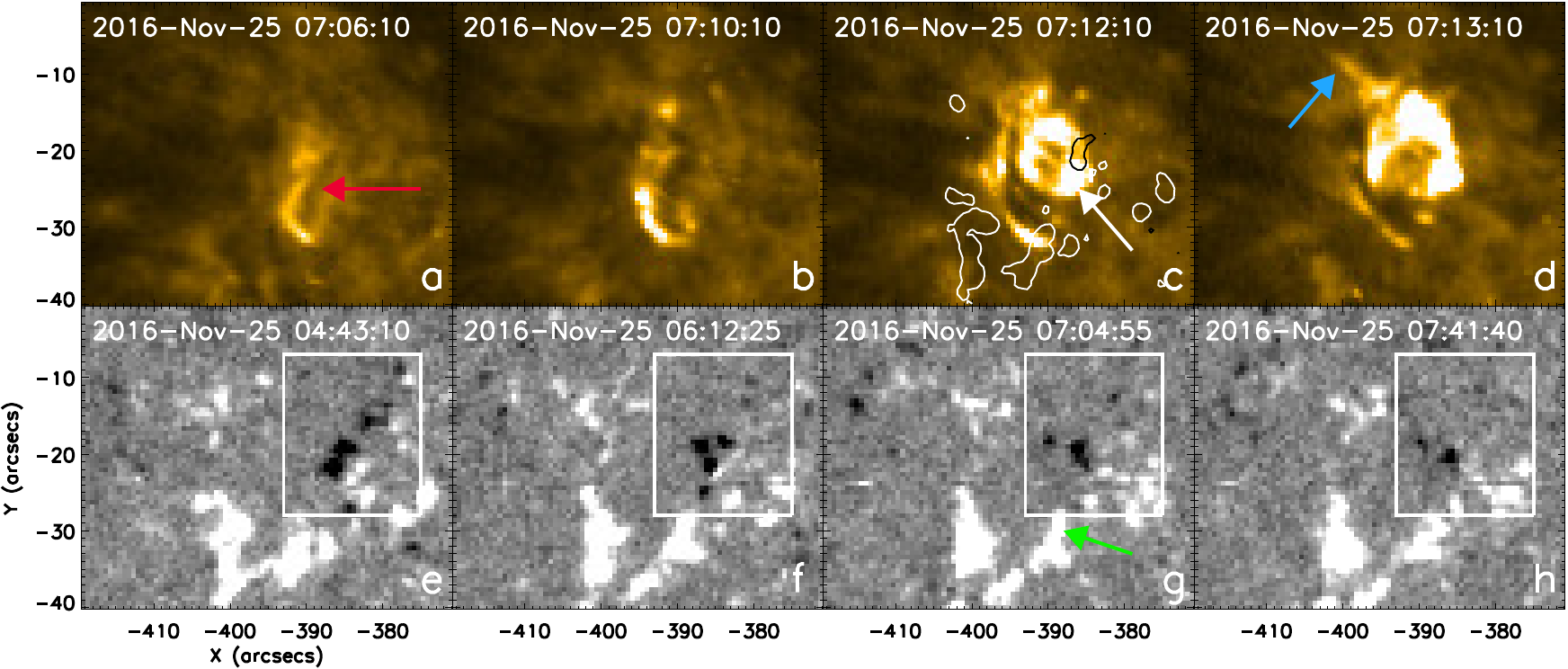}
	\caption{An example quiet region jet; observed on 2016 November 25  (Event 51). Panels (a)-(d) show 171 \AA{} AIA intensity images. The red arrow in panel (a) points to the erupting minifilament. The white arrow in panel (c) shows the JBP. The blue arrow in panel (d) indicates the jet spire next to the (mostly confined) erupting minifilament. Panels (e)-(h) show HMI magnetogram images of the same region. The green arrow in panel (g) points to the majority flux patch, to which EUV jet-base brightenings extend in (c). The boxed areas in panels (e)-(h) enclose the measured negative flux plotted in Figure \ref{fig:Nov25plot}. We have included movies in the online version of this paper showing the complete evolution of this jet in AIA 171 \AA{} (MOVIE5) and in the corresponding HMI magnetogram data (MOVIE6).}
	\label{fig:Nov25image}
\end{figure*}
A coronal hole jet was observed on March 22, 2017 (see Table \ref{tab:jetmeasurements} for measurements). Similar to the jet discussed in section \ref{sec:Jan2}, we observed a minifilament over the magnetic neutral line between the majority polarity flux patch (negative) and the minority polarity flux patch (positive, Fig. \ref{fig:Mar22image}). However, in this case, two jets occurred in close proximity from different magnetic neutral lines. We measured the larger jet, but shortly after it occurred, at 4:56 UT, a smaller jet occurred at 5:07 UT. We did not measure the smaller, second jet because its proximity to the larger jet made it difficult to obtain accurate measurements.
The pre-jet minifilament (preceding the first jet) was visible at least 1 hour prior to the jet spire occurring (MOVIE3, available in the online version of this paper). At $\sim$ 04:50 UT, the minifilament began to  rise (Fig. \ref{fig:Mar22image}(a)) and the JBP appeared (Figures \ref{fig:Mar22image}(b,f)) at the location (neutral line) of the minifilament  before the eruption. The growth duration of the spire was $15\pm2$ minutes. One minute prior to formation of the jet spire, we measured the jet base to be 8700 km wide. \par

Figure \ref{fig:Mar22plot} shows the measured positive minority flux from $\sim$ 5 hours prior to the jet at 04:56UT, to 30 minutes after jet onset. The measured positive flux is located in the jet base region indicated by the white box in Figure \ref{fig:Mar22image} panels (e)-(h). As with the jet discussed in section \ref{sec:Jan2}, on average there is an overall decrease in measured flux leading up to the jet (red line, Fig. \ref{fig:Mar22plot}) that occurs as the opposite-polarity flux patches approach the magnetic neutral line and cancel.\par 

\subsection{Quiet Region Jets} \label{sec:QR}
	We observed 30 EUV coronal jets in quiet regions from September 2016 to May 2017 (see table \ref{tab:jetmeasurements}). We present two examples of these quiet region jets in sections \ref{sec:Nov25} and \ref{sec:April10}.

\subsubsection{Quiet Region Jet on November 25, 2016 (Event 51)} \label{sec:Nov25}

A quiet region coronal jet was observed on November 25, 2016. In this jet region there are two jet-producing minifilament eruptions from the same neutral line: a partial minifilament eruption producing a weak jet at 06:21 UT (and there is an even weaker precursor to this jet at ~06:00 UT),	which we have not listed in Table \ref{tab:jetmeasurements}; and a complete minifilament eruption producing a final jet at 07:11 UT
 (MOVIE5, available in the online version of this paper). As with the previously discussed events, pre-jet cancellation occurred at the base of both of these jets, and the continued cancellation led to both jets. Similar homologous jets from the same neutral line have also been observed by \citet{Panesar2017} in their study of quiet region jets (and in \citet{Sterling2017} in active region jets). One minute prior to the final jet onset, the width of the jet base region was approximately 8100 km. A minifilament starts to rise at 07:01:22 UT (MOVIE5 and Fig. \ref{fig:Nov25image} (a)). At 07:12 UT we observed a JBP (Fig. \ref{fig:Nov25image} (c), white arrow) at the site of minifilament formation as well as an external brightening at a nearby patch of positive majority flux (Fig. \ref{fig:Nov25image}(c), green arrow in \ref{fig:Nov25image}(g)). After the rise of the minifilament, the jet spire started to extend and grew for $4\pm2$ minutes (Fig. \ref{fig:Nov25image} (d)). These observations follow behavior similar to that of the coronal hole jet discussed in section \ref{sec:Jan2}.  \par

\begin{figure}[h]
	\centering
	\plotone{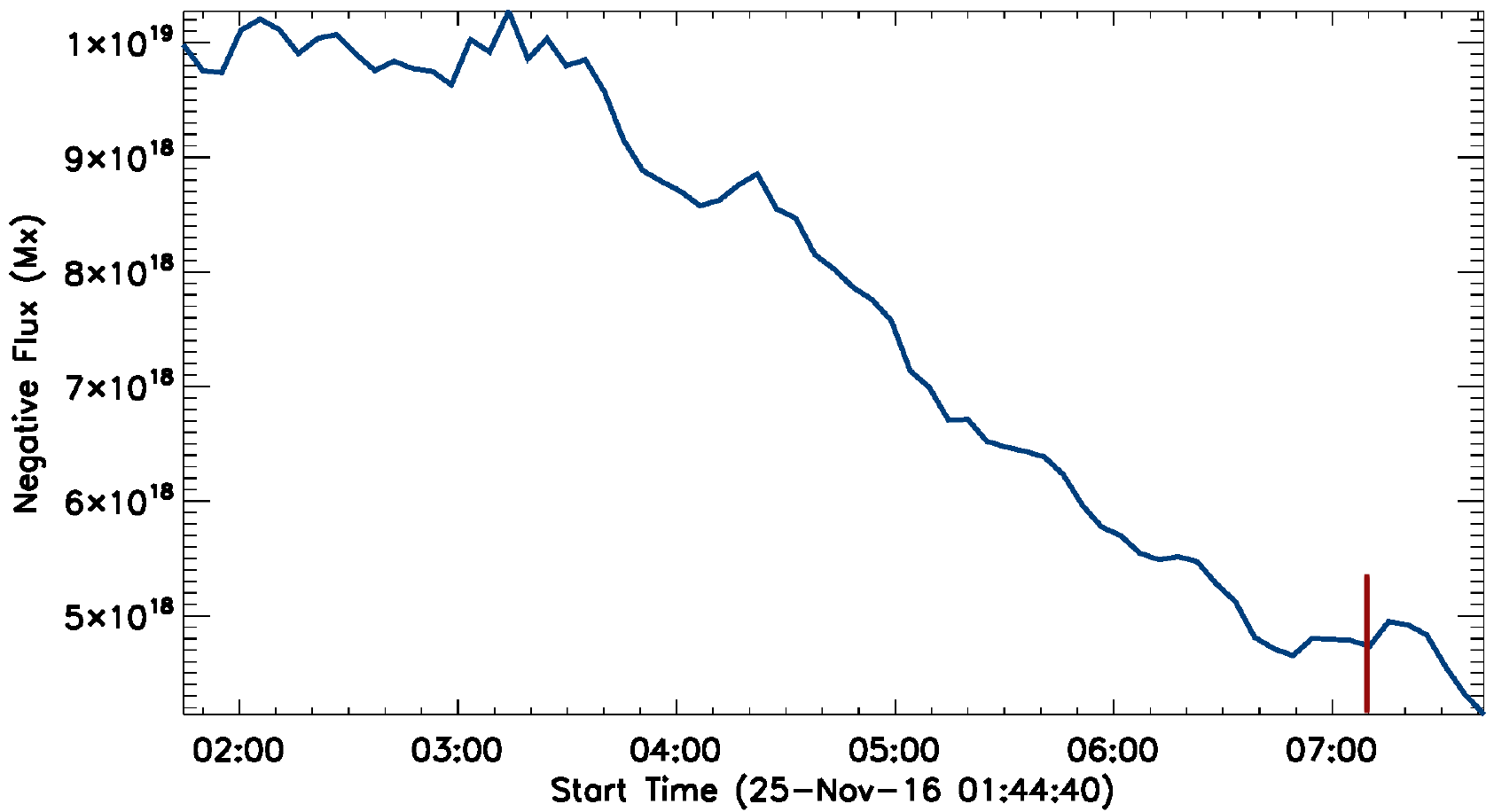}
	\caption{Flux-time plot from the base of the quiet region jet of Figure \ref{fig:Nov25image}. This shows the minority (negative) flux integrated over the box in Figure \ref{fig:Nov25image}(e)-(h) as a function of time. The red line designates the time of jet onset (07:09 UT).}
	\label{fig:Nov25plot}
\end{figure}
\begin{figure*}[t]
	\centering
	\includegraphics[width=6in]{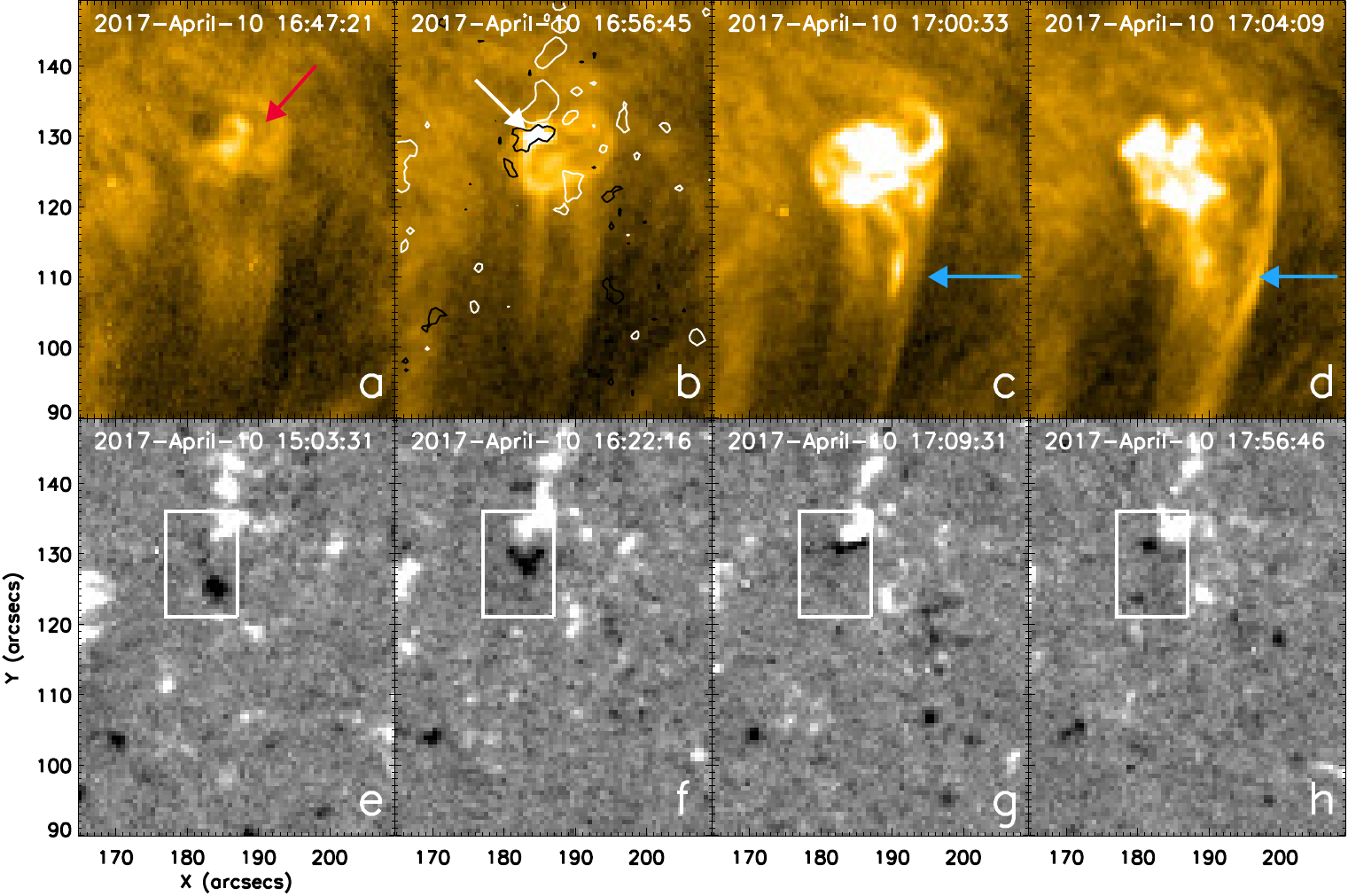}
	\caption{A second example of a quiet region jet; observed on 2017 April 10  (Event 5). Panels (a)-(d) show 171 \AA{} AIA intensity images. The red arrow in panel (a) points to the pre-eruption minifilament. The white (positive) and black (negative) contours in panel (b) represent the flux at the time of the AIA image. The white arrow in panel (b) shows the JBP. The blue arrows in panels (c and d) indicate the jet spire. Panels (e)-(h) show HMI magnetogram images of the same region. The boxed areas in panels (e)-(h) enclose the measured negative flux plotted in Figure \ref{fig:April10plot}. Movies showing the complete evolution of this jet in AIA 171 \AA{} (MOVIE7) as well as in the corresponding HMI magnetogram data (MOVIE8) are available in the online version of this paper.}
	\label{fig:April10image}
\end{figure*}
We measured the minority (negative polarity in this case) flux prior to and during the jet in the region indicated by the white box in Figure \ref{fig:Nov25image} panels (e)-(h). We present this measured flux as a function of time in Figure \ref{fig:Nov25plot}. Starting at 03:30 UT, about 4 hours prior to the jet, we see a continuous decrease in the overall negative flux in the jet base region leading up to the jet at 07:09 UT. Prior to this time, flux was coalescing, which caused the increase in measured flux at 3:00 UT (Fig. \ref{fig:Nov25plot}). Similar to the jets discussed in section \ref{sec:CH}, this decrease in flux occurs as opposite-polarity flux patches were approaching the magnetic neutral line and canceling with each other.

\subsubsection{Quiet Region Jet on April 10, 2017 (Event 5)} \label{sec:April10}
We observed a quiet region coronal jet at 16:56 UT on April 10, 2017 . As with the jets discussed in section \ref{sec:CH}, we observed a minifilament (Fig. \ref{fig:April10image} (a)) that lifted off from a magnetic neutral line, followed by the appearance of a JBP (Fig. \ref{fig:April10image} (b)) at that same location. During the eruption onset, the minifilament slowly rose to produce the jet spire (Fig. \ref{fig:April10image} (c)). One minute before the appearance of the jet spire, the jet base region was approximately 8300 km in diameter. Upon eruption of the minifilament, the jet spire grew for $10\pm1$ minutes. In the AIA 171 images we see some smaller flows starting at 16:39 UT that precede the main jet at 16:56 UT. These smaller flows are accompanied by brightenings in the jet base region, leading us to believe that they are homologous weaker jets resulting from the same magnetic neutral line (like those discussed in section \ref{sec:Nov25}).\par
\begin{figure}
	\centering
	\plotone{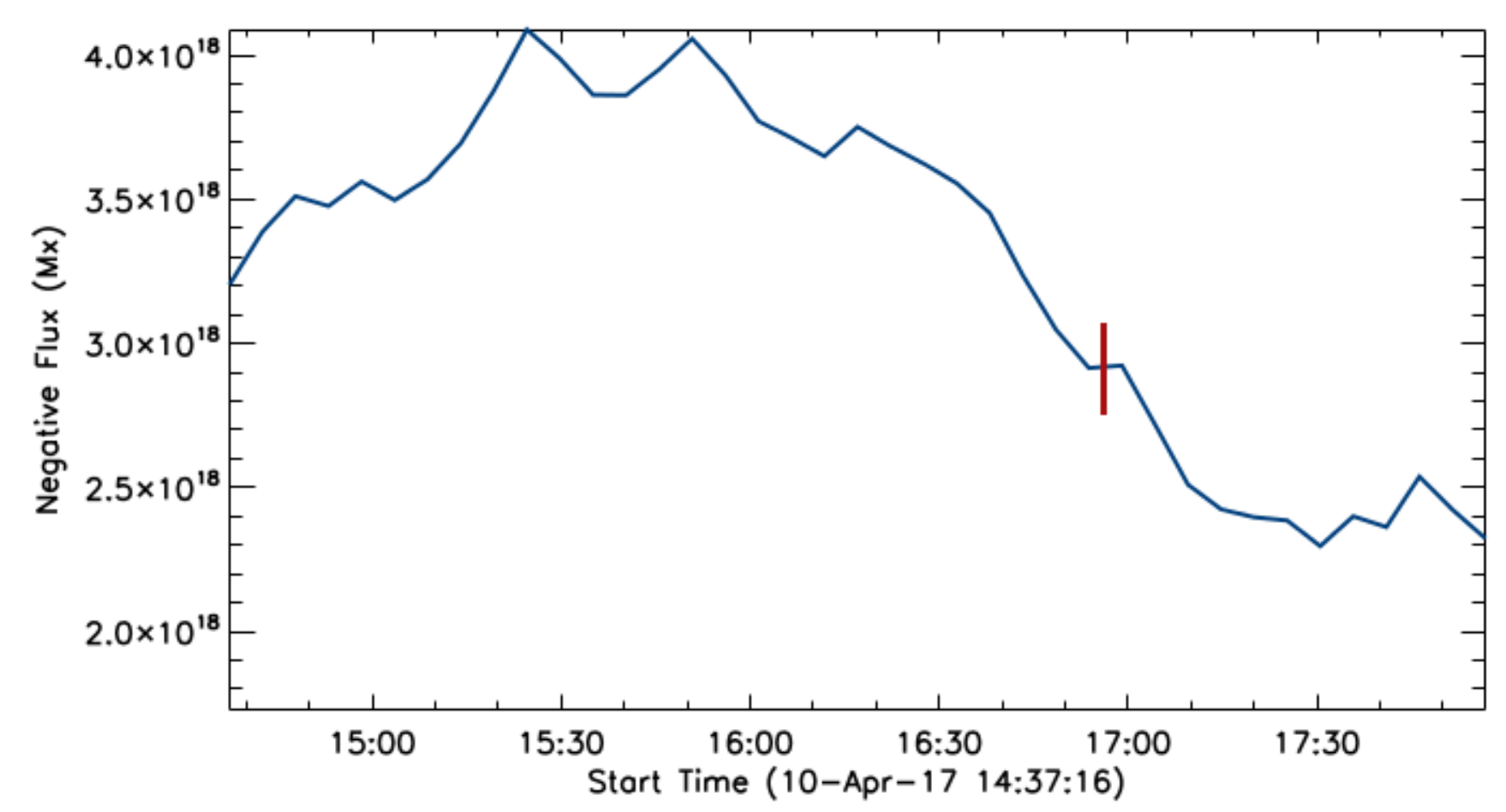}
	\caption{Flux-time plot from the base of the quiet region jet of Figure \ref{fig:April10image}. This shows the minority negative flux integrated over the box in Figure \ref{fig:April10image}(e)-(h) as a function of time. The red line designates the time of jet onset (16:56 UT).}
	\label{fig:April10plot}
\end{figure}

We measured the negative minority flux in the jet base region and plotted it as a function of time in Figure \ref{fig:April10plot}. The flux was measured in the region bounded by the white box in Figure \ref{fig:April10image}(e)-(h). Starting two hours prior to the jet, the positive and negative polarity flux patches approached the magnetic neutral line and canceled.  This behavior is reflected in the continuous overall decrease of negative flux that we see in Figure \ref{fig:April10plot} leading up to the jet at 16:56 UT. Prior to $\sim$15:30 UT, negative flux coalesced in the measured region, causing the measurable increase in flux visible at 15:30 UT in Fig. \ref{fig:April10plot}. \par

\subsection{Results for 60 jets}

We made a visual assessment of each jet, for (a) the behavior of the magnetic flux
changes in the time period near the time of jet onset, and (b) whether we could detect
a minifilament eruption at the time of the jet onset. \par

For most of the events, 51 in total (85\%), we observed clear convergence and apparent cancellation of magnetic flux occurring along with the jet.  There were only six quiet Sun jet cases, and three coronal hole jet cases, where we did not find clear evidence of flux cancellation; these nine exceptions are indicated in Table 1. In the nine cases where we did not observe clear flux cancellation, generally the flux was very weak at the location and time of the jet, so we consider that the measurement error was high. In some cases weak flux cancellation might be occurring; we expect that to have been the situation in several of these cases.  \par

If we consider an erupting minifilament to be dark (absorbing) material expelled with the jet when viewed in AIA 171\AA{}  movies, then we can identify erupting minifilaments in 56 events (all but four events), as indicated in Table 1. In one of the four exceptions (2016 Dec 3), the jet base is very
small, so if a minifilament exists, it may be too small for AIA to resolve. In the three remaining exceptions any minifilament is either too faint or absent.
Thus, while several of the visible minfilaments are relatively faint or otherwise tricky to detect (for example, the minifilament of 2017 Jan 3 (event 38) is difficult to recognize because it erupts toward the observer), we can identify them in over 90\% of the events (56/60).\par

Thus, our conclusion is that a large majority of jets in quiet Sun and coronal holes are produced by minifilament eruptions, and those eruptions are triggered to erupt by magnetic flux cancellation at the jet base region.


\section{Summary and Discussion}
	We have examined 60 solar coronal jets in quiet regions and coronal holes with EUV images from SDO/AIA and magnetogram data from SDO/HMI. In the large majority of instances, the jet was preceded by a minifilament eruption prepared and triggered by continuous flux cancellation at the magnetic neutral line within the jet base. This is consistent with the findings and proposed picture by \citet{Sterling2015} and \citet{Panesar2016}. Additionally, for our 60-jet sample we find a mean jet spire-growth duration of $9\pm4$ minutes and a mean jet base width of $8800\pm3100$km. \par
	
	Our jet spire-growth-duration measurements are very similar to the EUV jets in \citet{Panesar2016} ($\sim$ 12 minutes) and the X-ray jets of \citet{Savcheva2007} ($\sim$10 minutes). Our jet base width is nearly the same as the jet widths of \citealt{Savcheva2007} (8000 km) and the mean erupting-minifilament size of \citet{Sterling2015}. However, our mean jet base width is roughly half to that of \cite{Panesar2016} ($\sim$17,000 km)  plausibly because \cite{Panesar2016} found their jets from viewing JHelioviewer movies of full-disk AIA 171\AA\ coronal EUV images, whereas we found our jets by viewing JHelioviewer movies of zoomed-in partial disk AIA 171\AA\ coronal EUV images of quiet regions and coronal holes. That is, the \cite{Panesar2016} study may have preferentially selected jets of larger size scale compared to the jets selected in this study. \par

	The large majority of minifilament eruptions we observed are in agreement with \citet{Sterling2015} and \citet{Panesar2016}, as shown in Figure \ref{fig:jetfigRON}. In each case, prior to the jet, a minifilament resides between opposite polarity flux patches at the magnetic neutral line. As these flux patches approach the magnetic neutral line they continuously cancel and the minifilament eventually begins to erupt outwards. This eruption results in runaway internal reconnection within the erupting minifilament-carrying field as well as runaway external reconnection with the encountered outward-reaching field. The runaway internal reconnection creates the JBP that we observe in Figures \ref{fig:Jan2image}(b), \ref{fig:Mar22image}(c), \ref{fig:Nov25image}(c), and \ref{fig:April10image}(b). The runaway external reconnection produces new field lines, including the far-reaching field line along which the jet spire forms (Figures \ref{fig:Jan2image}(d), \ref{fig:Mar22image}(d), \ref{fig:Nov25image}(d), \ref{fig:April10image}(d)). The runaway external reconnection also produces external brightenings between the minority-polarity patch of the pair of canceling flux patches and a nearby patch of majority polarity flux (Figures \ref{fig:jetfigRON}(c), \ref{fig:Jan2image}(b), \ref{fig:Nov25image}(c)). 
	
    In conclusion, we report observations of 60 random on-disk coronal jets in quiet regions and coronal holes. From our observations, we find that at least for the great majority of these jets, minifilament eruption is the driver and magnetic flux cancellation is the trigger mechanism for coronal hole and quiet region coronal jets. \par

\section{Acknowledgements}
R.A.M gives special thanks to Dr. Navdeep Panesar for her time and mentorship throughout the research process, and to the NASA Marshall Space Flight Center, the Center for Space Plasma and Aeronomic Research at the University of Alabama in Huntsville, and the National Science foundation for supporting this study. R.A.M was suported by the Research Experience for Undergraduates (REU) program at UAH and MSFC over summer 2017, which was funded by the National Science Foundation under grant No.~AGS-1460767. N.K.P's research was supported by an appointment to NPP at the NASA/MSFC, administered by USRA under contract with NASA. A.C.S and R.L.M acknowledge the support from the NASA HGI program, and by the Hinode Project. We acknowledge the use of SDO data.\par

\bibliographystyle{apj}
\bibliography{jetbib}

\end{document}